\documentclass[aps,prc,10pt,showpacs,showkeys,twocolumn,superscriptaddress,groupedaddress]{revtex4-1}

\usepackage[english]{babel}
\usepackage{indentfirst}
\usepackage{amssymb}
\usepackage{braket}
\usepackage{bm}
\usepackage{amsxtra}
\usepackage{amsmath}
\usepackage{supertabular}
\usepackage{multirow}
\usepackage{color}
\usepackage [mathcal]{eucal}
\usepackage{graphicx}
\usepackage{graphics}
\usepackage{lipsum}
\usepackage{dsfont}
\usepackage{mathtools}
\usepackage{comment}

\def\id{\mathds{1}}

     \def\chiral4lo{$\mathrm{N}^4\mathrm{LO}$}

\begin{document}


\title{Towards a microscopic description of nucleus-nucleus collisions}

\author{Matteo Vorabbi$^{1}$}
\author{Michael Gennari$^{2,3}$}
\author{Paolo Finelli$^{4}$}
\author{Carlotta Giusti$^{5}$}
\author{Petr Navr\'atil$^{2,3}$}

\affiliation{$~^{1}$ School of Mathematics and Physics, University of Surrey, Guildford, GU2 7XH, United Kingdom}

\affiliation{$~^{2}$TRIUMF, 4004 Wesbrook Mall, Vancouver, British Columbia, V6T 2A3, Canada
}

\affiliation{$~^{3}$University of Victoria, 3800 Finnerty Road, Victoria, British Columbia V8P 5C2, Canada
}

\affiliation{$~^{4}$Dipartimento di Fisica e Astronomia, 
Universit\`{a} degli Studi di Bologna and \\
INFN, Sezione di Bologna, Via Irnerio 46, I-40126 Bologna, Italy
}

\affiliation{$~^{5}$INFN, Sezione di Pavia,  Via A. Bassi 6, I-27100 Pavia, Italy
}

\date{\today}


\begin{abstract} 
We present the first results of a comprehensive microscopic approach to describe nucleus-nucleus elastic collisions by means of an optical potential derived at
first order in multiple-scattering theory and computed by folding the projectile and target nuclear densities with the nucleon-nucleon $t$ matrix, which describes the interaction between
each nucleon of the projectile and each nucleon of the target.
Chiral interactions are consistently used in the calculation of the $t$ matrix and of the nonlocal nuclear densities, which are computed within the
ab initio no-core shell model.
Cross sections calculated for $\alpha$ collisions on $^{12}$C and $^{16}$O at projectile energies in the range 100-300 MeV are presented and compared with available
data.
For momentum transfer $q$ up to about $1.0$ fm$^{-1}$ our results are in good agreement with the experimental data, whereas for higher momenta a reduction of the imaginary contributions is needed.
\end{abstract}


\maketitle


Nucleus-nucleus collisions are fundamental processes that provide insight into the properties of nuclear matter and the dynamics of heavy-ion interactions.
They are usually described by traditional phenomenological models which often rely on tunable parameters to fit experimental data, limiting their predictive
power, particularly concerning exotic nuclei.
The study of microscopic optical potentials (OPs) has been and still is a fascinating area of research in nuclear physics, since they play a crucial role in understanding the
interactions of nucleons in atomic nuclei \cite{FOLDY1969447,PEHodgson_1971,DICKHOFF2019252,Hebborn_2023} and in describing nucleon-nucleus and
nucleus-nucleus interactions. From the experimental point of view, in the past decade a large effort has been made to study short-lived exotic nuclei using proton elastic
scattering in inverse kinematics at facilities such as the CSRe storage ring of HIRFL-CSR \cite{2002NIMPA.488...11X}, GSI/FAIR \cite{kalantar2018nustar} and the RIBF at
RIKEN \cite{2009AIPC.1120..241S}. In the near future the dynamics of elastic scattering involving light bound nuclei will likely receive considerable attention as a probe
of unexpected deviations from the theoretical expectations based on our present knowledge of the nucleon-nucleon interaction.

In general, OPs have broad implications across various nuclear reactions, such as nucleon-induced reactions or fusion processes. OPs find applications in astrophysics,
where they are crucial for modelling nucleosynthesis in stars and understanding stellar evolution \cite{10.3389/fspas.2020.00009}, but they also play a significant role in
nuclear energy applications, such as reactor design and fuel cycle simulations \cite{annurev:/content/journals/10.1146/annurev-nucl-101918-023708}.

The importance of a microscopic description of nucleus-nucleus collisions is also connected to the relevance for hadron-therapy applications and space radiation protection \cite{PhysRevC.81.024613}. In fact, modelling the space radiation environment is a crucial step in planning for space missions and the analysis of
shielding efficiency.
The space radiation transport codes used to describe the transport of ions and secondary particles produced from nuclear collisions require knowledge of the relevant
nuclear reactions, among them elastic scattering. At the moment such reactions are described by phenomenological models \cite{10.3389/fphy.2021.788253,
PhysRevC.81.045805} due to theoretical limitations and the scarcity of data.

Since microscopic approaches are derived from first principles, they ensure consistency with the underlying physics. These methods incorporate fundamental interactions,
nuclear structure information, and experimental constraints to construct a comprehensive picture of nuclear systems. As a result, microscopic approaches provide a more
fundamental understanding of the phenomena and avoid ad hoc assumptions usually included in phenomenological descriptions. Microscopic models offer the advantage of
an enhanced predictive power, allowing them to rigorously extrapolate beyond the experimental data used to construct the OP.
This predictive capability is crucial for investigating unexplored regions of the nuclear chart, exotic nuclei, and nuclear reactions under extreme conditions, e.g., those
encountered in astrophysical environments. By employing solid theoretical models it is possible to make significant progress in the interpretation of the complex nature of
these potentials and their impact on nuclear phenomena \cite{Hebborn_2023}.


In this manuscript we present a model to derive a microscopic nucleus-nucleus OP from the multiple-scattering theory. We refer to the
Supplemental Material \cite{SupMat} for a sketch of the main steps of the derivation of the OP, as well as for some computational and numerical details.
At first order in the theory, corresponding to the single-scattering approximation, the OP for elastic nucleus-nucleus collisions is obtained from the double-folding
integral of the free nucleon-nucleon ($NN$) $t$ matrix and the densities of the projectile ($\mathbb{P}$) and target ($\mathbb{T}$) nuclei,
$\rho^{(\mathbb{P})}$ and $\rho^{(\mathbb{T})}$, respectively.
In our approach, the $NN$ $t$ matrix plays the role of an effective interaction between a proton or neutron ($p$ or $n$) in $\mathbb{P}$ and a
proton or neutron in $\mathbb{T}$. 
The general formula of our OP is then given by~\cite{SupMat}
\onecolumngrid
\begin{equation}\label{nucleus_opticalpotworkeq}
\begin{split}
U_{\mathrm{el}} ({\bm q} , {\bm K};E) &= \sum_{\alpha, \beta =p,n} \int d {\bm Q}_{\mathbb{P}} \int d {\bm Q}_{\mathbb{T}} \; \eta ({\bm q} , {\bm K} , {\bm Q}_{\mathbb{P}} , {\bm Q}_{\mathbb{T}} ) \,
t_{\alpha \beta} \left[ {\bm q} , \frac{1}{2} \left( \frac{A_{\mathbb{P}}+A_{\mathbb{T}}}{A_{\mathbb{P}} A_{\mathbb{T}}} {\bm K} - \sqrt{\frac{A_{\mathbb{P}}-1}{A_{\mathbb{P}}}} {\bm Q}_{\mathbb{P}} + \sqrt{\frac{A_{\mathbb{T}}-1}{A_{\mathbb{T}}}} {\bm Q}_{\mathbb{T}} \right) ; \mathcal{E} \right] \\
&\times \rho_{\alpha}^{(\mathbb{P})} \left( {\bm Q}_{\mathbb{P}} - \frac{1}{2} \sqrt{\frac{A_{\mathbb{P}}-1}{A_{\mathbb{P}}}} {\bm q} , {\bm Q}_{\mathbb{P}} + \frac{1}{2} \sqrt{\frac{A_{\mathbb{P}}-1}{A_{\mathbb{P}}}} {\bm q} \right) 
\times \rho_{\beta}^{(\mathbb{T})} \left( {\bm Q}_{\mathbb{T}} + \frac{1}{2} \sqrt{\frac{A_{\mathbb{T}}-1}{A_{\mathbb{T}}}} {\bm q} , {\bm Q}_{\mathbb{T}} - \frac{1}{2} \sqrt{\frac{A_{\mathbb{T}}-1}{A_{\mathbb{T}}}} {\bm q} \right) \, .
\end{split}
\end{equation}
\twocolumngrid

In Eq.~(\ref{nucleus_opticalpotworkeq}) the variables ${\bm q} = {\bm k}^{\prime} - {\bm k}$  and $2 {\bm K} = {\bm k}^{\prime} + {\bm k}$ are the momentum transfer
and the average momentum, respectively, where ${\bm k}$ and ${\bm k}^{\prime}$ are the initial and final relative momenta in the nucleus-nucleus frame;
${\bm Q}_{\mathbb{P}}$ and ${\bm Q}_{\mathbb{T}}$ are integration variables, $A_{\mathbb{P}}$ and $A_{\mathbb{T}}$ the number of nucleons of the projectile and of
the target and $\eta$ the M\o ller factor that imposes the Lorentz invariance between different reference systems, as discussed in the Supplemental Material \cite{SupMat}.
\nocite{Watson:1957zz,SATCHLER1979183,cucinotta1988theory,PhysRevC.36.1839,Werneth:2014cta,Werneth:NASA,PhysRevC.24.1400,austern1970direct,satchler1983direct,feshbach1992theoretical,PhysRevC.103.024604,PhysRevC.52.1992,PhysRevC.57.1378,PhysRevC.102.024616,Girlanda2011,Kravvaris2023,PhysRevC.109.065501,Weppner_dissertation,osti_4002515,Navratil:2009ut,Barrett:2013nh,PhysRevC.96.024004,Navratil:2007zn,PhysRevC.101.014318,Gysbers2019,Bogner:2003wn,Bogner:2009bt,PhysRevC.97.014601}

The energy $\mathcal{E}$ at which the $NN$ $t$ matrix is computed is a complicated function of ${\bm K}$, ${\bm Q}_{\mathbb{P}}$, and
${\bm Q}_{\mathbb{T}}$, and makes the calculation of Eq.(\ref{nucleus_opticalpotworkeq}) unfeasible. In the present work this is set to one-half the kinetic
energy of a single nucleon in the projectile nucleus (see Ref. \cite{SupMat} for more details).

We note that the $t$ matrix entering Eq.~(\ref{nucleus_opticalpotworkeq}) only contains the central (spin-independent) term, because the other terms lead to small or
vanishing contributions. Consequently, our OP only contains the central term. However, we would like to stress that this term contains both real ($V$) and imaginary ($W$) parts, such that $U = V + i W$. In particular, the imaginary part is produced in the calculation by the imaginary part of the $t$ matrix.

The use of the free two-nucleon scattering operator is an approximation that reduces the complexity of the
original many-body problem to a form in which we only have to solve two-body equations, neglecting the effect of nuclear binding on the two interacting nucleons.

In this work we present numerical results of the cross sections obtained with the microscopic OP of Eq.~(\ref{nucleus_opticalpotworkeq}) for a selected set of cases
of $^{4}$He elastic scattering off $^{12}$C and $^{16}$O at three incoming energies, 104, 130, and 240 MeV, for which experimental data are available.

For the calculation of the nonlocal density of the projectile and target nuclei in momentum space we employ the no-core shell model (NCSM)
approach~\cite{Navratil:2009ut, Barrett:2013nh}. The NCSM is based on an expansion of the nuclear wave function in a harmonic oscillator basis 
characterized by the frequency $\hbar \omega$ and the basis truncation parameter $N_{max}$, which specifies the number of nucleon excitations above the lowest
energy configuration allowed by the Pauli principle.

All our calculations have been performed with the $NN$ chiral interaction developed by Entem {\it et al.} \cite{PhysRevC.96.024004} up to the fifth order (N$^{4}$LO) with a
500 MeV cutoff and the three-nucleon ($3N$) local-nonlocal chiral interaction at N$^{2}$LO presented in Refs. \cite{Navratil:2007zn,PhysRevC.101.014318},
with the low-energy constants fixed at $c_D = -1.8$ and $c_E =-0.31$ \cite{Gysbers2019}.
The same $NN$ chiral interaction used to compute the nuclear densities is consistently used in the calculation of the $NN$ $t$ matrix.
For the three nuclei we employed $\hbar \omega = 18$ MeV and a $\lambda_{\mathrm{SRG}} = 1.8$ fm$^{-1}$ cutoff for the similarity renormalization
group (SRG) \cite{Bogner:2003wn, Bogner:2009bt} procedure (including the SRG induced 3N force in all the calculations), and
we performed calculations up to $N_{max} = 8$ for $^{12}$C and $^{16}$O, and $N_{max} = 16$ for $^{4}$He.


\begin{figure}[t]
\begin{center}
\includegraphics[scale=0.35]{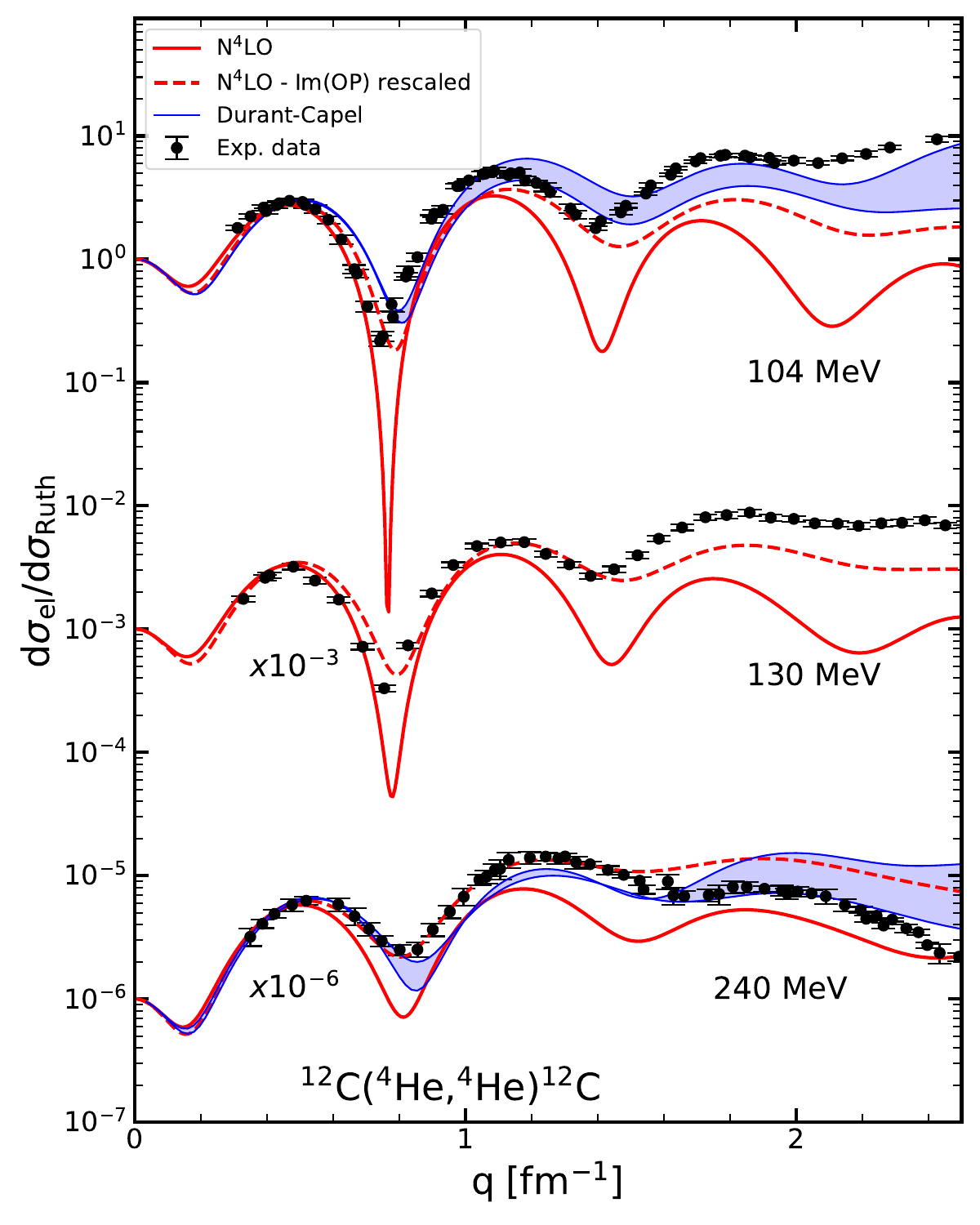}
\caption{
Ratio of the differential cross section to the Rutherford cross section as a function of the transferred momentum $q$ for the reaction $^{12}$C($^{4}$He,$^{4}$He)$^{12}$C. 
Calculations are performed at projectile energies $E = 104, 130$, and $240$ MeV. Experimental data \cite{PhysRevC.97.014601,HAUSER196981,PhysRevC.68.014305}
are shown by black circles with corresponding error bars. 
The solid red lines are obtained using our microscopic OP of
Eq.(\ref{nucleus_opticalpotworkeq}), red dashed lines are obtained rescaling the imaginary part of our OP by a factor 0.5. The results obtained in Ref. \cite{Durant:2020pgr} are shown by the shaded blue bands for a comparison.
\label{4He_12C} }
\end{center}
\end{figure}

\begin{figure}[t]
\begin{center}
\includegraphics[scale=0.35]{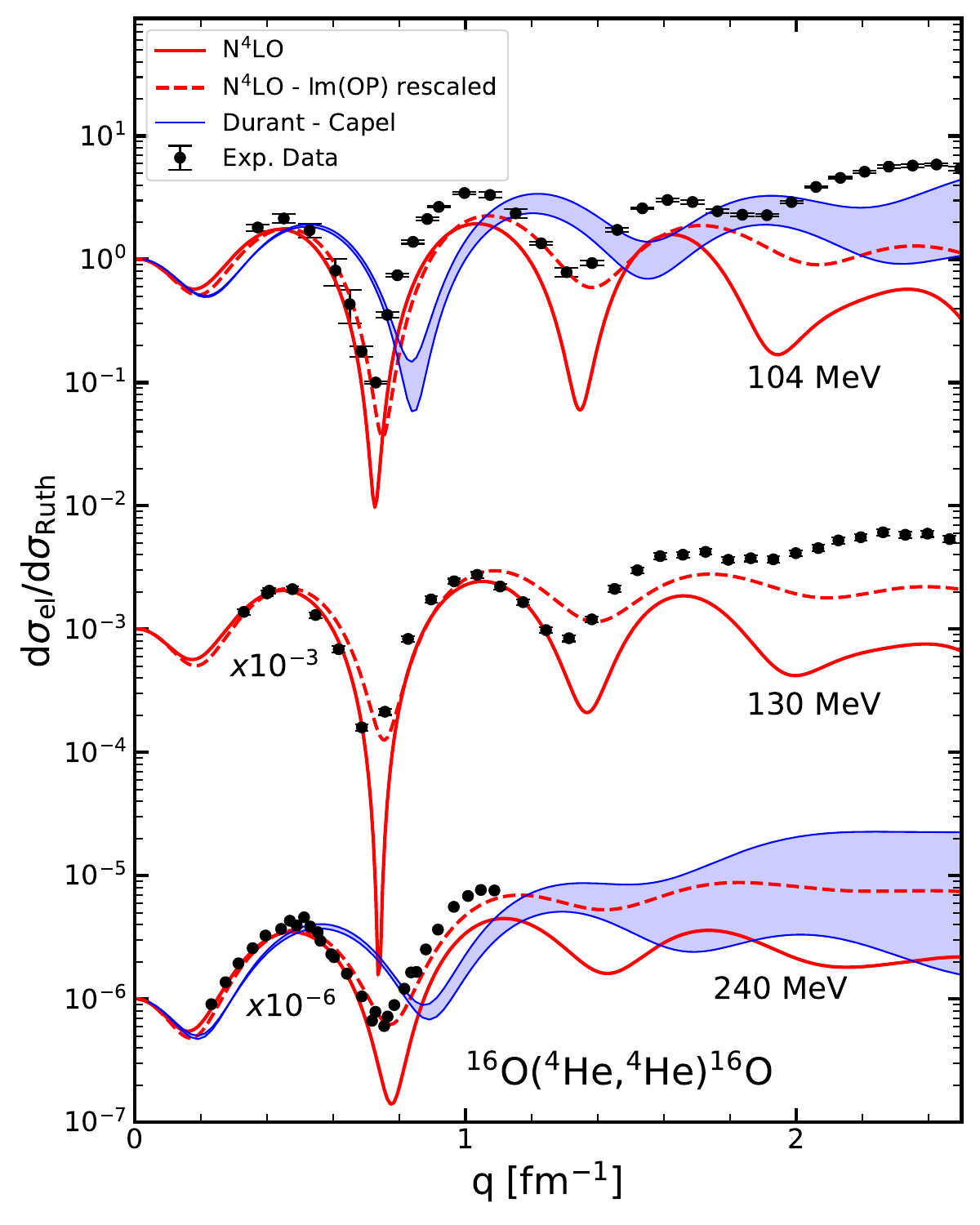}
\caption{The same as in Fig. \ref{4He_12C} but for the reaction $^{16}$O($^{4}$He,$^{4}$He)$^{16}$O.
\label{4He_16O} }
\end{center}
\end{figure}
In Figs.~\ref{4He_12C} and \ref{4He_16O} we display in solid red lines the differential cross sections (divided by the Rutherford cross section) obtained
with our OP as a function of the momentum transfer $q$ for the $^{12}$C($^4$He,$^4$He)$^{12}$C and $^{16}$O($^4$He,$^4$He)$^{16}$O reactions, respectively,
at the three aforementioned energies.
In both figures our OP gives a good description (size, shape, position of the minima) of the experimental data for values of $q \leq 1.0$ fm$^{-1}$, showing that our
microscopic model, which does not contain adjustable parameters, is theoretically well-founded and should have good predictive power in situations for which empirical
data are not yet available, e.g., for exotic nuclei.
In both figures, however, for larger values of $q$ our results underpredict the data and the disagreement increases with $q$.
The reason for the disagreement is not clear, but it is likely due to the approximate description of the interaction of a nucleon in the projectile and a nucleon in the target
by the free $NN$ $t$ matrix, which seems to introduce an excessive contribution from the absorptive imaginary part of the OP in the model. 
In order to test this hypothesis, we performed calculations with an artificially reduced imaginary component  multiplied by 0.5.
This reduction (the corresponding results are depicted by red dashed lines in the figures) barely changes the behaviour at small $q$, whereas it substantially improves the agreement up to large values of the momentum transfer.

As a benchmark, recent calculations by Durant and Capel \cite{Durant:2020pgr} are also shown in the figures by the light blue bands.
In Ref. \cite{Durant:2020pgr} the OP is obtained from the double-folding method using chiral $NN$ interactions at N$^2$LO \cite{DURANT2018668} and realistic nuclear
densities; the bands give theoretical uncertainties by use of different cutoff radii in the chiral interaction and different nuclear densities.
Few years ago we performed a similar investigation for proton-nucleus elastic scattering, showing that a satisfactory level of convergence could be achieved once
chiral interactions developed up to N$^4$LO are employed \cite{Vorabbi:2017rvk}. For the nucleus-nucleus case, the sensitivity of our theoretical predictions upon different interactions and truncation schemes has been studied in the Supplemental Material \cite{SupMat},  as shown in Figs. S1 and S2.

In Ref. \cite{Durant:2020pgr} the imaginary part of the OP is obtained
from the real one, either, as in phenomenological approaches \cite{ALVAREZ200393,PEREIRA2009330}, multiplying the real part by a proportionality constant
\begin{equation}\label{phenomenological_prescription}
W = N_W V \, ,
\end{equation}
with $N_W$ in the range 0.5-0.8, or linking the two parts by applying the Kramers-Kronig dispersion relations
\begin{equation}\label{dispersion_relation}
W(E) = - \frac{1}{\pi} \, \mathcal{P} \int_{-\infty}^{\infty} \frac{V_{\mathrm Ex}(E^{\prime})}{E^{\prime} - E} \, dE^{\prime} \, ,
\end{equation}
where $\mathcal{P}$ denotes the Cauchy principal value of the integral and $V_{\mathrm Ex}$ the exchange part of the real component.
Eq.(\ref{dispersion_relation}) ensures that the OP is constrained by the correct energy dependence of the imaginary absorptive part. In Ref. \cite{Durant:2020pgr} the two prescriptions give similar results, even if the Kramers-Kronig relations yield better agreement with data, for large values of $q$ and without any free parameter.
The light blue bands in Figs.~\ref{4He_12C} and \ref{4He_16O} are computed with the imaginary part obtained with Eq.(\ref{dispersion_relation}). In any case both prescriptions give results closer to our dashed line than to our solid lines, i.e., to our results with the reduced imaginary part of our OP.

\begin{figure}[t]
\begin{center}
\includegraphics[scale=0.4]{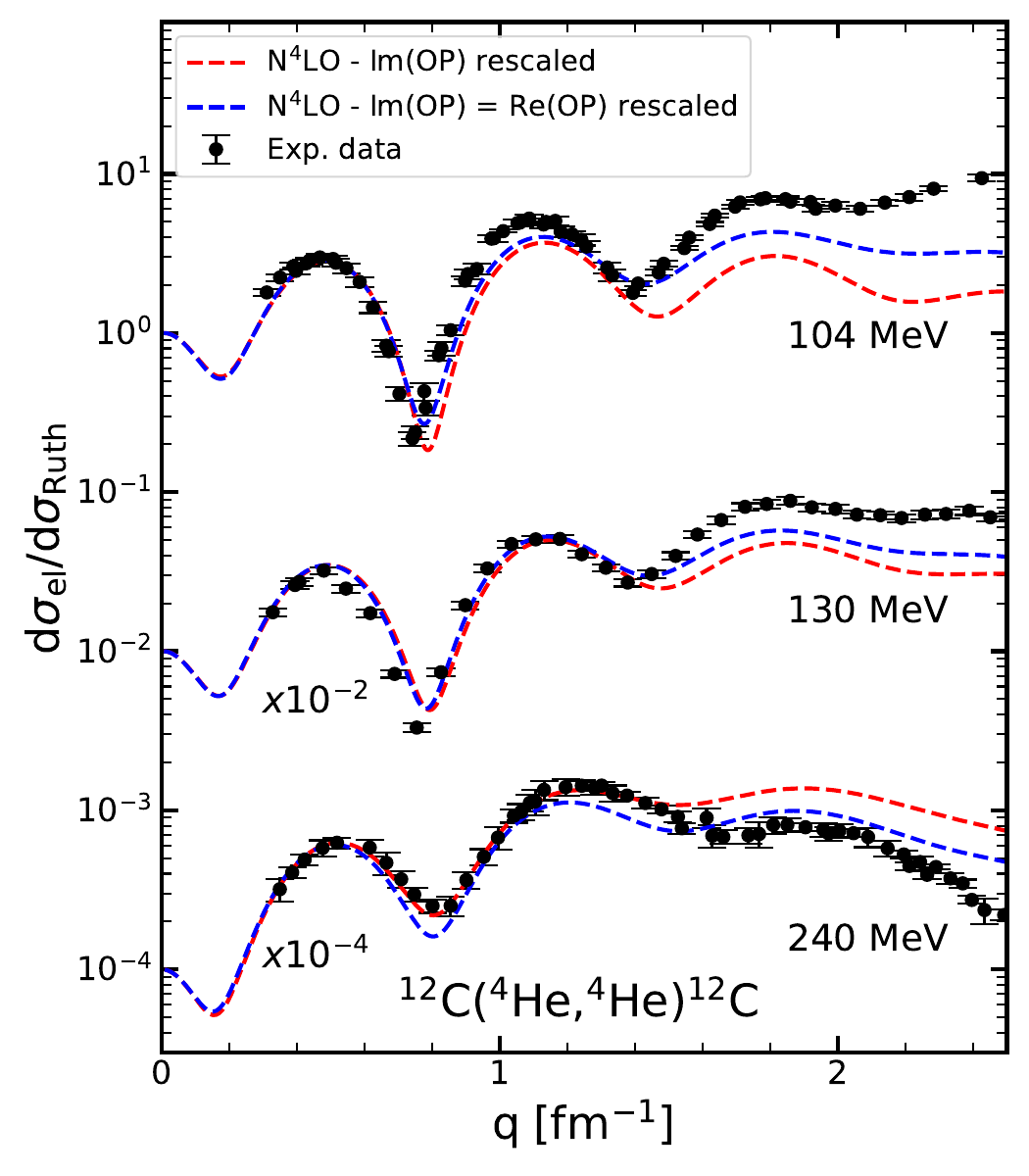}
\caption{ The dashed red lines are the same as in Fig.~\ref{4He_12C}, the dashed blue lines are the corresponding results obtained replacing the imaginary part of our OP with its real part rescaled by the same factor $0.5$.
\label{real_potential_rescaling} }
\end{center}
\end{figure}

Additional information concerning the connection between the phenomenological and the microscopic description of the absorptive component of the optical potential can be obtained by looking separately at the real and imaginary parts.
As mentioned prior, in phenomenological approaches a typical prescription is given by Eq.(\ref{phenomenological_prescription}).
This choice is based on the underlying assumption that $W$ and $V$ have a similar behaviour as functions of the relevant coordinates.
In Fig.~\ref{real_potential_rescaling} we show with the dashed blue lines the results obtained adopting Eq.(\ref{phenomenological_prescription}),
where the imaginary part of our microscopic OP is replaced by the real part multiplied by 0.5. We can see
that for all three energies the results obtained with the phenomenological prescription are very close to those obtained by
rescaling the imaginary part (dashed red lines). In our analysis we have plotted the real and imaginary
parts and have confirmed that they have very similar shapes. The reason why the two curves are not exactly the same is because
the real and imaginary parts of the OP have slightly different depths and in this work we always rescaled the two parts with the
same factor.

\begin{figure}[t]
\begin{center}
\includegraphics[scale=0.4]{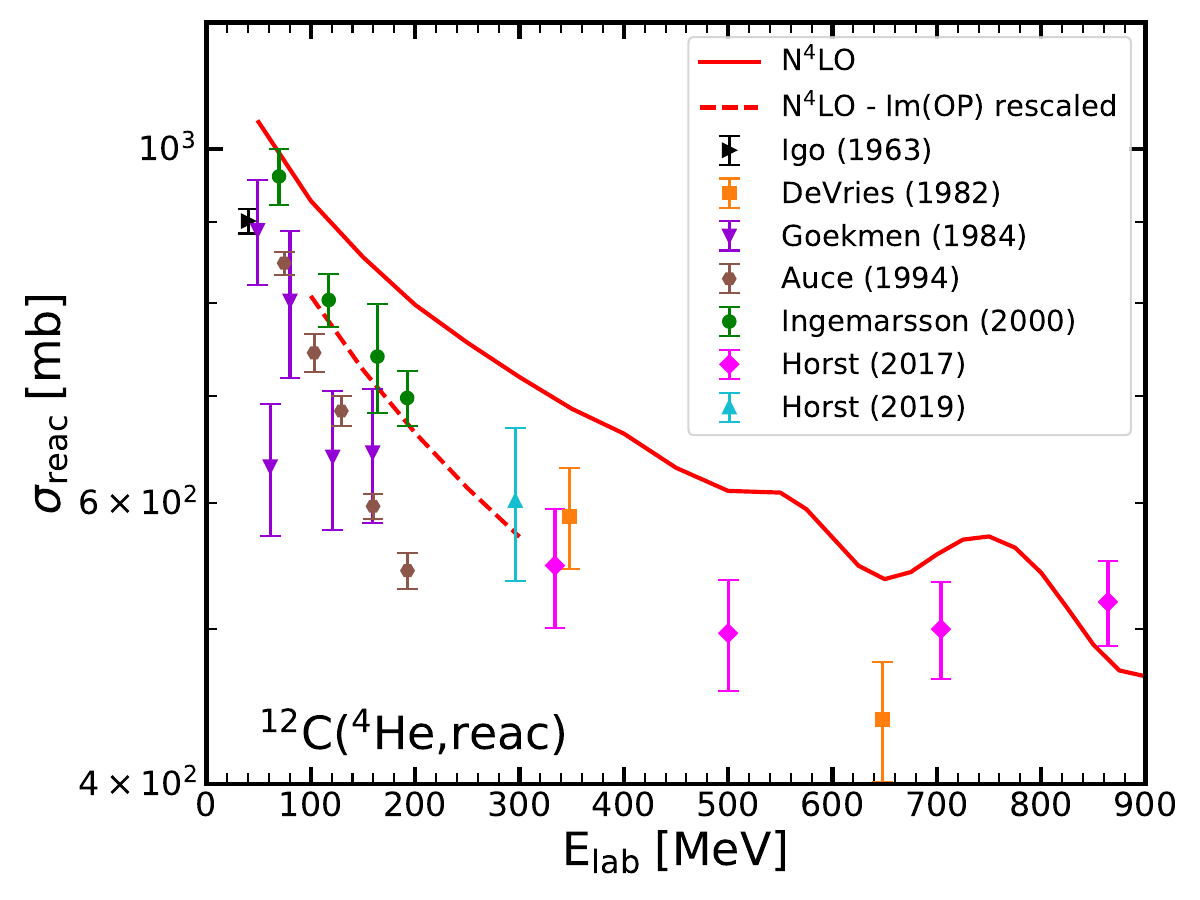}
\caption{Reaction cross section $^{12}$C($^{4}$He, reac) as a function of the projectile laboratory energy $E_{lab}$. The solid red line is the result of our microscopic OP,
the dashed red line is obtained rescaling the imaginary part of our OP by a factor 0.5. Experimental data are shown by filled coloured symbols with the corresponding error 
bars: rightward black triangles \cite{Wilkins:1963zza}, orange squares \cite{DeVries:1982zz}, downward violet triangles \cite{Gokmen:1984ukj}, brown circles \cite{Abele:1994ufy, Ingemarsson:1994zz, Auce:1994zz}, green circles \cite{Ingemarsson:2000vfz}, and pink diamonds \cite{Horst:2017rnm}.
\label{reaction_cross_section} }
\end{center}
\end{figure}

In Fig.~\ref{reaction_cross_section} we show the reaction cross section $^{12}$C($^{4}$He, reac) in comparison with experimental
data \cite{Wilkins:1963zza,DeVries:1982zz,Abele:1994ufy, Ingemarsson:1994zz, Auce:1994zz,Ingemarsson:2000vfz,Horst:2017rnm} that cover a large energy range
(10-1100 MeV). Our results overestimate data but are able to describe their overall behaviour, which is not trivial since microscopic approaches are usually
able to reproduce angular distributions but not necessarily integrated quantities. Once the imaginary part has been halved, in the energy range 100-300 MeV (in which we
are more confident about the approximations of our model and where the available differential cross section data allow the rescaling procedure) our results displayed by the dashed red line fall close to the experimental data.


In conclusion, we have presented the first results of a microscopic approach to describe nucleus-nucleus collisions.
Our microscopic OP provides a good description of the experimental data for values of momentum transfer up to 1.0 fm$^{-1}$ and underestimates the data for
larger values. The disagreement between theoretical and experimental results in this region indicates that the OP seems to be too absorptive and a simple reduction
of the imaginary part seems to confirm that.

The theoretical reason for this excess of absorption is however not clear. 
The model contains several approximations which might explain the disagreement with the data,
such as the use of the free $NN$ $t$ matrix to describe the interaction between a nucleon in the projectile and one in the target, the prescription used to fix the
energy $\mathcal{E}$ at which the $t$ matrix is computed, or the single-scattering approximation adopted to derive the OP. 
Including multiple scattering effects can play a significant role, as discussed in Ref. \cite{Crespo:1992zz} for the nucleon-nucleus case. The inclusion of medium effects \cite{Arellano:1995zz} could be also helpful to reduce the disagreement with experimental data.

In spite of all the adopted approximations and without any free parameters our results are able to describe the overall behaviour of the reaction cross section,
which is not trivial, and are in remarkable agreement with the experimental differential cross section for values of $q \leq$ 1.0 fm$^{-1}$.
In our opinion, this is a clear indication that these first results represent a significant step towards a microscopic description and a more fundamental understanding
of nuclear collisions.

Despite the overall quality of the description of experimental observables, our approach is well suited for improvements. Our theoretical model, we believe, can be improved through two main aspects. On the one hand, by analyzing and studying the various prescriptions for the energy variable of the propagator; on the other hand, by conducting a thorough investigation of the imaginary component of the optical potential, particularly exploring the potential existence of a connection with the dispersive relations approach.

Our results highlight the significance of incorporating quantum mechanical effects (through sophisticated many-body methods like multiple-scattering theory) and detailed
nuclear descriptions of projectiles and targets (by ab initio methods like the NCSM approach) to accurately describe the optical potential.
These findings not only provide deeper insights into the fundamental processes governing nuclear interactions but also pave the way for more accurate and predictive
models in nuclear physics, particularly for exotic nuclei. Future work should aim to refine these models further and explore their implications for a broader
range of nuclear phenomena.



We would like to thank P. Capel and V. Durant  (University of Mainz) for useful discussions and providing results presented in Ref. \cite{Durant:2020pgr}.

This work used the DiRAC Data Intensive service (DIaL3) at the University of Leicester, managed by the University of Leicester Research Computing Service on behalf of the STFC DiRAC HPC Facility (www.dirac.ac.uk). The DiRAC service at Leicester was funded by BEIS, UKRI and STFC capital funding and STFC operations grants. DiRAC is part of the UKRI Digital Research Infrastructure. This work used the DiRAC Complexity system, operated by the University of Leicester IT Services, which forms part of the STFC DiRAC HPC Facility (www.dirac.ac.uk ). This equipment is funded by BIS National E-Infrastructure capital grant ST/K000373/1 and STFC DiRAC Operations grant ST/K0003259/1. DiRAC is part of the National e-Infrastructure.

M.G. and P.N. acknowledge support from the NSERC Grant No. SAPIN-2022-00019. TRIUMF receives federal funding via a contribution agreement with the National Research Council of Canada.
Computing support also came from an INCITE Award on the Summit and Frontier super- computers of the Oak Ridge Leadership Computing Facility (OLCF) at ORNL and from the Digital Research Alliance of Canada.














\clearpage

\onecolumngrid

\begin{center}
{\bf \large Supplemental Material}
\end{center}

\twocolumngrid

\renewcommand{\thefigure}{S\arabic{figure}}
\renewcommand{\thetable}{S\arabic{table}}
\renewcommand{\theequation}{S\arabic{equation}}
\renewcommand{\thepage}{\arabic{page}}
\setcounter{figure}{0}
\setcounter{table}{0}
\setcounter{equation}{0}


\section{Introduction}

In this Supplemental Material we sketch the derivation of the nucleus-nucleus potential adopted in the paper entitled {\it Towards a microscopic description of nucleus-nucleus collisions}  and derived using the framework and techniques of the multiple-scattering theory \cite{Watson:1957zz}.
By considering each nucleon-nucleon ($NN$) interaction as an individual scattering event and aggregating the effects of these multiple scatterings, we can construct a comprehensive picture of the entire collision process by the introduction of an optical potential (OP).


Inspired by the theoretical approaches derived in Refs. \cite{SATCHLER1979183, cucinotta1988theory, PhysRevC.36.1839, Werneth:2014cta, Werneth:NASA} we present in the following the theoretical formalism we propose for a microscopic description, in the momentum-space representation, of nucleus-nucleus collisions based on the multiple-scattering approach.
With the purpose of extending the well-known formalism of the spectator expansion within the multiple-scattering theory to the nucleus-nucleus elastic scattering 
case, we  consider the general case of a collision between a projectile nucleus $\mathbb{P}$ composed of $A_{\mathbb{P}}$ nucleons and a target nucleus ${\mathbb T}$ of $A_{\mathbb{T}}$ nucleons. We assume that the nucleons belonging to $\mathbb{P}$ are distinguishable from the ${\mathbb T}$ nucleons: issues concerning fermionic antisymmetrization are extensively discussed in Ref. \cite{PhysRevC.24.1400}. At the moment, we ignore effects of Pauli antisymmetrization between projectile and target nucleons with the assumption that  these effects are supposed to be small for small scattering angles, i.e. in the region where the first-order optical potential can be safely trusted.

\section{Theoretical Framework}

\subsection{Optical potential within the multiple scattering theory}

At the basis of our theoretical derivation we employ the Lippmann-Schwinger (LS) equation. In fact, the LS equation is used to describe the scattering transition amplitudes that are an essential ingredient to derive all relevant empirical observables such as total/differential cross section or polarization dependent quantities, e.g. the analyzing power. In case of elastic transitions it is customary to write a set of equivalent relations: one for the definition of the optical potential and the other one for the proper calculation of the elastic scattering process.

Therefore, as stated above, our starting point for the nucleus-nucleus ($\mathbb{P} \mathbb{T}$) transition operator $T$ is the many-body LS equation, that is conventionally considered the exact treatment of a scattering problem, as extensively illustrated in Refs. \cite{austern1970direct, satchler1983direct, feshbach1992theoretical}
\begin{equation}\label{nucleus_generalscatteq}
T = V + V G_0 (E) T \, ,
\end{equation}
where the operator $V$ represents the external interaction between the two nuclei, such that the Hamiltonian for the entire $\mathbb{P} \mathbb{T}$ system is given by
\begin{equation}
H_{\mathbb{P} \mathbb{T}} = H_0 + V \, ,
\end{equation}
where
\begin{equation}
H_0 = K_0 + H_{\mathbb{P}} + H_{\mathbb{T}} \, .
\end{equation}
Here $K_0$ is the kinetic energy operator for the projectile-target relative motion and $H_{\mathbb{P}}$ and $H_{\mathbb{T}}$ are the internal Hamiltonians of the projectile
and target nuclei, respectively. Asymptotically, the system is in an eigenstate of $H_0$, and the free (many-body) propagator $G_0 (E)$ for the projectile-target system is
\begin{equation}
G_0 (E) = \frac{1}{E - H_0 + i \epsilon} \, .
\end{equation}
Now we introduce the projection operators $P$ and $Q$ that satisfy the relation
\begin{equation}
P + Q = \id \, ,
\end{equation}
where $P$ is conventionally taken to project onto the elastic channel.
Thus the projector $P$ can be defined as
\begin{equation}
P = \ket{\Phi_{\mathbb{P}}^{(0)} \, \Phi_{\mathbb{T}}^{(0)}} \bra{\Phi_{\mathbb{P}}^{(0)} \, \Phi_{\mathbb{T}}^{(0)}} \, ,
\end{equation}
where $\ket{\Phi_{\mathbb{P}}^{(0)} \, \Phi_{\mathbb{T}}^{(0)}} \equiv \ket{\Phi_{\mathbb{P}}^{(0)}} \ket{\Phi_{\mathbb{T}}^{(0)}}$, with $\ket{\Phi_{\mathbb{P}}^{(0)}}$ as the ground state of the projectile
nucleus and $\ket{\Phi_{\mathbb{T}}^{(0)}}$ as the ground state of the target nucleus. They must fulfil
\begin{align}
H_{\mathbb{P}} \ket{\Phi_{\mathbb{P}}^{(0)}} &= E_{\mathbb{P}}^{(0)} \ket{\Phi_{\mathbb{P}}^{(0)}} \, , \\
H_{\mathbb{T}} \ket{\Phi_{\mathbb{T}}^{(0)}} &= E_{\mathbb{T}}^{(0)} \ket{\Phi_{\mathbb{T}}^{(0)}} \, ,
\end{align}
where $E_{\mathbb{P}}^{(0)}$ and $E_{\mathbb{T}}^{(0)} $ are the ground-state energies.
Using the operators $P$ and $Q$ we can split Eq.(\ref{nucleus_generalscatteq}) into two parts, i.e., an integral equation for $T$
\begin{equation}\label{nucleus_firsttamp}
T = U + U G_0 (E) P T \, ,
\end{equation}
where $U$ is the OP operator, and an integral equation for $U$
\begin{equation}\label{nucleus_optpoteq}
U = V + V G_0 (E) Q U \, .
\end{equation}
With these definitions the transition operator for elastic scattering may be defined as $T_{\mathrm{el}} = PTP$, in which case Eq.~(\ref{nucleus_firsttamp}) can be written as
\begin{equation}\label{nucleus_elastictransition}
T_{\mathrm{el}} = P U P + P U P G_0 (E) T_{\mathrm{el}} \, .
\end{equation}
Thus the transition operator for elastic scattering is given by the knowledge of the operator $P U P$. The following theoretical treatment consists of a formulation of the
many-body equation (\ref{nucleus_optpoteq}), where expressions for $U$ are derived such that $P U P$ can be calculated without having to solve the
complete many-body problem. Here we only assume the presence of two-body forces, but this assumption could be easily relaxed, as shown
in Ref. \cite{PhysRevC.103.024604} in which chiral three-body terms have been approximated and treated as two-body density-dependent contributions. Within this framework, the external interaction $V$ can be written as
\begin{equation}
V = \sum_{i=1}^{A_{\mathbb{P}}} \sum_{j=A_{\mathbb{P}}+1}^{A_{\mathbb{P}}+A_{\mathbb{T}}} v_{ij} \, ,
\end{equation}
where the indices $i$ and $j$ belong to the nucleons in the projectile and target, respectively, and the two-body potential $v_{ij}$ acts between the $i$th nucleon in the projectile nucleus and the $j$th nucleon in the target nucleus. In a similar way, the operator $U$ for the optical potential can be expressed as
\begin{equation}
U = \sum_{i=1}^{A_{\mathbb{P}}} \sum_{j=A_{\mathbb{P}}+1}^{A_{\mathbb{P}}+A_{\mathbb{T}}} U_{ij} \, ,
\end{equation}
where $U_{ij}$ is given by the recursive relation
\begin{equation}\label{nucleus_optpoteq2}
U_{ij} = v_{ij} + v_{ij} G_0 (E) Q \sum_{k=1}^{A_{\mathbb{P}}} \sum_{l=A_{\mathbb{P}}+1}^{A_{\mathbb{P}}+A_{\mathbb{T}}} U_{kl} \, .
\end{equation}
Through the introduction of an effective scattering operator $\tau_{ij}$ between the nucleons $i$ and $j$, which satisfies
\begin{equation}\label{nucleus_firstordertau}
\tau_{ij} = v_{ij} + v_{ij} G_0 (E) Q \tau_{ij} \, ,
\end{equation}
and still possesses a many-body nature thanks to the propagator $G_0 (E)$, we can rearrange Eq.~(\ref{nucleus_optpoteq2}) as
\begin{equation}
U_{ij} = \tau_{ij} + \tau_{ij} G_0 (E) Q \sum_{\substack{k=1 \\ k\neq i}}^{A_{\mathbb{P}}} \sum_{\substack{l=A_{\mathbb{P}}+1 \\ l\neq j}}^{A_{\mathbb{P}}+A_{\mathbb{T}}} U_{kl} \, .
\end{equation}
This rearrangement process can continue for all the $A_{\mathbb{P}}$ projectile nucleons and $A_{\mathbb{T}}$ target nucleons, leading to a generalized version of
the spectator expansion \cite{PhysRevC.52.1992} for the optical potential operator.
Since we are exploring this approach for the first time, in this work we are only interested in the first term of this series and thus we can approximate the expression for the OP operator with its leading order as
\begin{equation}\label{nucleus_firstorder}
U \simeq \sum_{i=1}^{A_{\mathbb{P}}} \sum_{j=A_{\mathbb{P}}+1}^{A_{\mathbb{P}}+A_{\mathbb{T}}} \tau_{ij} \, .
\end{equation}
At this point we introduce the additional approximation of replacing the operator $\tau_{ij}$ with the free $NN$ $t$ matrix operator $t_{i j}$ describing the
interaction of the two nucleons in free space and satisfying
\begin{equation}\label{pt_t_matrix}
t_{ij} = v_{ij} + v_{ij} \, g_{ij} \, t_{ij} \, ,
\end{equation}
where
\begin{equation}\label{pt_nn_propagator}
g_{i j} = \left[ E - h_i^{(\mathbb{P})} - h_j^{(\mathbb{T})} + i \epsilon \right]^{-1}
\end{equation}
is the free $NN$ propagator.
Under this approximation, the final expression of the optical-potential operator is given by
\begin{equation}\label{nucleus_firstorder_ia}
U = \sum_{i=1}^{A_{\mathbb{P}}} \sum_{j=A_\mathbb{P}+1}^{A_{\mathbb{P}}+A_{\mathbb{T}}} t_{ij} \, .
\end{equation}
We notice that even if for nucleon-nucleus elastic scattering this approximation (also known as impulse approximation) has been widely used at intermediate
scattering energies, in the present case the situation is more complicated because the projectile nucleon is not free anymore and it is bound in the projectile nucleus. 

\subsection{Double folding optical potential}

Now we introduce the basis
\begin{equation}
\ket{\Phi_{\mathbb{P}}^{(0)} \, \Phi_{\mathbb{T}}^{(0)} \, {\bm k}} \equiv \ket{\Phi_{\mathbb{P}}^{(0)}} \ket{\Phi_{\mathbb{T}}^{(0)}} \ket{\bm k}
\end{equation}
to project the $T_{\mathrm{el}}$ operator of Eq.(\ref{nucleus_elastictransition}), which is evaluated in the $\mathbb{P} \mathbb{T}$ frame.
With this basis the equation for the elastic transition amplitude becomes
\begin{equation}\label{nucleus_mastertransitioneq}
T_{\mathrm{el}} ({\bm k}^{\prime} , {\bm k}) = U_{\mathrm{el}} ({\bm k}^{\prime} , {\bm k}) + \int d {\bm k}^{\prime \prime} \frac{U_{\mathrm{el}} ({\bm k}^{\prime} , {\bm k}^{\prime \prime}) T_{\mathrm{el}} ({\bm k}^{\prime \prime} , {\bm k})}{E - E (k^{\prime \prime}) + i \epsilon} \, ,
\end{equation}
which can be solved using standard techniques.
Starting from the definition $U_{\mathrm{el}} \equiv P U P$ and
expressing $U$ with Eq.(\ref{nucleus_firstorder_ia}), now we have to derive an explicit expression for the elastic OP.
This can be done along the same line followed for the derivation of the nucleon-nucleus OP \cite{PhysRevC.52.1992}.
We simply assume two active nucleons, one in the projectile nucleus (nucleon $A$) and another one in the target nucleus (nucleon $B$) and we evaluate
the following matrix element
\begin{equation}\label{general_t_matrix_element}
\braket{t_{AB}} \equiv \braket{\Phi_{\mathbb{T}}^{(0)} \, \Phi_{\mathbb{P}}^{(0)} | t_{A B} (E) | \Phi_{\mathbb{P}}^{(0)} \, \Phi_{\mathbb{T}}^{(0)} } \, ,
\end{equation}
where $t_{AB}$ represents the $NN$ $t$ matrix operator of Eq.(\ref{pt_t_matrix}) describing the interaction between nucleon $A$ and nucleon $B$.
Since the analysis of Eq.(\ref{general_t_matrix_element}) is the same for all nucleons, the final OP is simply obtained as the sum of Eq.(\ref{general_t_matrix_element})
for each possible choice of $A$ and $B$, i.e., proton-proton, proton-neutron, neutron-proton, and neutron-neutron.
The evaluation of Eq.(\ref{general_t_matrix_element}) is done working in the $\mathbb{P} \mathbb{T}$ frame, inserting decompositions of the identity, and using the
definition of the ground-state (translationally invariant) one-body densities for the projectile and target nuclei.
This leads to the final expression of the OP given by
\onecolumngrid
\begin{equation}\label{nucleus_opticalpotworkeq_sm}
\begin{split}
U_{\mathrm{el}} ({\bm q} , {\bm K};E) &= \sum_{\alpha, \beta =p,n} \int d {\bm Q}_{\mathbb{P}} \int d {\bm Q}_{\mathbb{T}} \; \eta ({\bm q} , {\bm K} , {\bm Q}_{\mathbb{P}} , {\bm Q}_{\mathbb{T}} ) \,
t_{\alpha \beta} \left[ {\bm q} , \frac{1}{2} \left( \frac{A_{\mathbb{P}}+A_{\mathbb{T}}}{A_{\mathbb{P}} A_{\mathbb{T}}} {\bm K} - \sqrt{\frac{A_{\mathbb{P}}-1}{A_{\mathbb{P}}}} {\bm Q}_{\mathbb{P}} + \sqrt{\frac{A_{\mathbb{T}}-1}{A_{\mathbb{T}}}} {\bm Q}_{\mathbb{T}} \right) ; \mathcal{E} \right] \\
&\times \rho_{\alpha}^{(\mathbb{P})} \left( {\bm Q}_{\mathbb{P}} - \frac{1}{2} \sqrt{\frac{A_{\mathbb{P}}-1}{A_{\mathbb{P}}}} {\bm q} , {\bm Q}_{\mathbb{P}} + \frac{1}{2} \sqrt{\frac{A_{\mathbb{P}}-1}{A_{\mathbb{P}}}} {\bm q} \right) 
\times \rho_{\beta}^{(\mathbb{T})} \left( {\bm Q}_{\mathbb{T}} + \frac{1}{2} \sqrt{\frac{A_{\mathbb{T}}-1}{A_{\mathbb{T}}}} {\bm q} , {\bm Q}_{\mathbb{T}} - \frac{1}{2} \sqrt{\frac{A_{\mathbb{T}}-1}{A_{\mathbb{T}}}} {\bm q} \right) \, .
\end{split}
\end{equation}
\twocolumngrid
In the previous expression, the vectors ${\bm Q}_{\mathbb{P}}$ and ${\bm Q}_{\mathbb{T}}$ are integration variables while
the vectors ${\bm q}$ and ${\bm K}$ are the momentum transfer and the average momentum, respectively. These are related to the
initial and final relative momenta ${\bm k}$ and ${\bm k}^{\prime}$ in the $\mathbb{P}\mathbb{T}$ frame through
\begin{align}
{\bm q} &= {\bm k}^{\prime} - {\bm k} \, , \\
{\bm K} &= \frac{1}{2} \big( {\bm k}^{\prime} + {\bm k} \big) \, .
\end{align}
Since we work in the $\mathbb{P} \mathbb{T}$ frame, the $t$ matrix appearing in Eq.(\ref{nucleus_opticalpotworkeq_sm}) needs to be defined in this reference frame,
so it must be transformed to the $\mathbb{P} \mathbb{T}$ frame through the M\o ller factor $\eta$, basically imposing $t_{\mathbb{P} \mathbb{T}} = \eta \, t_{NN}$, that is formally defined in the following.

\subsection{The M\o ller factor}

The M\o ller factor arises when it is necessary to perform a transformation from the ${\mathbb{P}} {\mathbb{T}}$ system to the $NN$ system in which the scattering amplitude is calculated from the microscopic chiral interactions. Our treatment is nothing else that a straightforward generalization of the conventional derivation introduced in Refs. \cite{Weppner_dissertation, osti_4002515} for the $NA$ case. The proof is rather simple since it is based on the requirement of the Lorentz-invariance of the reference systems transformations because the $\tau$ matrices previously introduced are not Lorentz-invariant by construction. We essentially impose
\begin{equation}
t_{\mathbb{P} \mathbb{T}} = \eta \, t_{NN} \, .
\end{equation} 
After some algebraic mnipulations, the M\o ller factor can be defined as follows
\begin{equation}\label{moellerfactor}
\begin{split}
\eta ({\bm q},&{\bm K},{\bm Q}_{\mathbb{P}} , {\bm Q}_{\mathbb{T}} ) \\
&= \frac{\sqrt{ E_N ({\bm \kappa}) E_N (- {\bm \kappa}) E_N ({\bm \kappa}^{\prime}) E_N (- {\bm \kappa}^{\prime}) }}{\sqrt{E_N ({\bm k}_{A_{\mathbb{P}}}) E_N ({\mathbf k}_{A_{\mathbb{T}}}) E_N ({\bm k}_{A_{\mathbb{P}}}^{\prime}) E_N ({\mathbf k}_{A_{\mathbb{T}}}^{\prime})}} \, ,
\end{split}
\end{equation}
where in general the energy terms are given by $E_N ({\bm k}) \equiv \sqrt{k^2 + m_N^2}$, where $m_N$ is the nucleon mass, and ${\bm k}$ is its 
momentum computed in a specific frame. The M\o ller factor (\ref{moellerfactor}) is by construction a real quantity and, as a consequence, does not contribute to the dynamical generation of the imaginary component of the Optical Potential (\ref{nucleus_opticalpotworkeq_sm}).

In this particular case the expressions for ${\bm \kappa}$ and ${\bm \kappa}^{\prime}$ (the initial and final relative momenta in the $NN$ frame, respectively)
that appear in Eq. (\ref{moellerfactor}) are given by
\begin{align}\label{nucleus_nn_momenta}
{\bm \kappa}^{\prime} &= \frac{1}{2} \big( a {\bm K} - b {\bm Q}_{\mathbb{P}} + c {\bm Q}_{\mathbb{T}} + {\bm q} \big) \, , \\
{\bm \kappa} &= \frac{1}{2} \big( a {\bm K} - b {\bm Q}_{\mathbb{P}} + c {\bm Q}_{\mathbb{T}} - {\bm q} \big) \, ,
\end{align}
where we defined
\begin{align}
a &= \frac{A_{\mathbb{P}} + A_{\mathbb{T}}}{A_{\mathbb{P}} A_{\mathbb{T}}} \, , \\
b &= \sqrt{\frac{A_{\mathbb{P}}-1}{A_{\mathbb{P}}}} \, , \\
c &= \sqrt{\frac{A_{\mathbb{T}}-1}{A_{\mathbb{T}}}} \, .
\end{align}
For the other momenta in the $\mathbb{P} \mathbb{T}$ frame we have
\begin{align}
{\bm k}_{A_{\mathbb{P}}}^{\prime} &= \frac{{\bm K}}{A_{\mathbb{P}}} - \sqrt{\frac{A_{\mathbb{P}} - 1}{A_{\mathbb{P}}}} {\bm Q}_{\mathbb{P}} + \frac{{\bm q}}{2} \, , \\
{\bm k}_{A_{\mathbb{P}}} &= \frac{{\bm K}}{A_{\mathbb{P}}} - \sqrt{\frac{A_{\mathbb{P}} - 1}{A_{\mathbb{P}}}} {\bm Q}_{\mathbb{P}} - \frac{{\bm q}}{2} \label{nucl_momentum_a_in_the_pt_frame} \, , \\
{\bm k}_{A_{\mathbb{T}}}^{\prime} &= - \frac{{\bm K}}{A_{\mathbb{T}}} - \sqrt{\frac{A_{\mathbb{T}} - 1}{A_{\mathbb{T}}}} {\bm Q}_{\mathbb{T}} - \frac{{\bm q}}{2} \, , \\
{\bm k}_{A_{\mathbb{T}}} &= - \frac{{\bm K}}{A_{\mathbb{T}}} - \sqrt{\frac{A_{\mathbb{T}} - 1}{A_{\mathbb{T}}}} {\bm Q}_{\mathbb{T}} + \frac{{\bm q}}{2} \label{nucl_momentum_b_in_the_pt_frame} \, ,
\end{align}
where ${\bm k}_{A_{\mathbb{P}}}$ and ${\bm k}_{A_{\mathbb{P}}}^{\prime}$ are the initial and final momenta of the $A_{\mathbb{P}}$-th nucleon in the projectile nucleus, and ${\bm k}_{A_{\mathbb{T}}}$ and ${\bm k}_{A_{\mathbb{T}}}^{\prime}$ are the initial and final momenta of the $A_{\mathbb{T}}$-th nucleon in the target nucleus.

As a first application we are mainly interested into nuclear collisions between spin-saturated nuclei, so, in this work we only consider the central part of the $NN$
$t$ matrix entering Eq.(\ref{nucleus_opticalpotworkeq_sm}).

\subsection{Energy prescription}

The energy $E$ in the left-hand side of Eq.(\ref{nucleus_opticalpotworkeq_sm}) represents the kinetic energy of the projectile nucleus in the laboratory frame. To calculate the OP with Eq.(\ref{nucleus_opticalpotworkeq_sm}) we need to calculate the total energy $E_{\mathbb{P} \mathbb{T}}$ in the $\mathbb{P} \mathbb{T}$ frame, that is just one term of the effective energy $\mathcal{E}$ appearing in the $t$ matrix. In fact, it is possible to express $\mathcal{E}$ 
in terms of the variables ${\bm K}$, ${\bm Q}_{\mathbb{P}}$, and ${\bm Q}_{\mathbb{T}}$,
obtaining \cite{PhysRevC.57.1378}
\begin{equation}\label{energy_expression}
\mathcal{E} = E_{\mathbb{P} \mathbb{T}}
- \frac{\left( \frac{A_{\mathbb{T}} - A_{\mathbb{P}}}{A_{\mathbb{P}} A_{\mathbb{T}}} {\bm K}
- \sqrt{\frac{A_{\mathbb{P}} - 1}{A_{\mathbb{P}}}} {\bm Q}_{\mathbb{P}} - \sqrt{\frac{A_{\mathbb{T}} - 1}{A_{\mathbb{T}}}} {\bm Q}_{\mathbb{T}} \right)^2}{4 m_N} \, ,
\end{equation}
where $m_N$ is the nucleon mass.
This shows the complicated energy dependence of the $t$ matrix that in principle should be taken into account during the evaluation of the double folding integral.
This would require the pre-computation of the $t$ matrix at many different energies, including negative energies, and an additional energy interpolation
during the calculation of Eq.(\ref{nucleus_opticalpotworkeq_sm}). Currently, this is computationally too expensive and cannot be achieved, so we need to introduce an approximation to simplify the numerical calculation of Eq.(\ref{nucleus_opticalpotworkeq_sm}).
A reasonable guess for the calculations we aim to perform is using the fixed-beam energy approximation.
This consists in evaluating the $t$ matrix at one half the kinetic energy available per nucleon (in the projectile nucleus) in the laboratory frame (which is equivalent to the on-shell center of mass energy for the two nucleons)
\begin{equation}\label{energy_approximation}
\mathcal{E} = \frac{1}{2} \frac{k_{lab}^2}{2m_N} \, .
\end{equation}
Here $k_{lab}$ represents the momentum of a single nucleon inside the projectile nucleus in the laboratory frame.
The approximation introduced through Eq.(\ref{energy_approximation}) is a forward scattering approximation which has been widely used in the context of nucleon-nucleus OP producing good results, but in
the present context it has never been tested. The advantage of this approximation is that it requires only one calculation of the $t$ matrix at the energy
given by Eq.(\ref{energy_approximation}), avoiding an additional energy interpolation of the $NN$ $t$ matrix.

\section{Numerical results}

In the following we present numerical results of the elastic cross sections obtained with the OP of Eq.(\ref{nucleus_opticalpotworkeq_sm}) for $^4$He elastic scattering
off $^{12}$C at the $^4$He laboratory energy of 130 MeV. Our results are also compared with the available experimental data.
The goal of these calculations is to test the stability of our results against different choices of $NN$ interactions used in the calculation of the nonlocal density and the
$NN$ $t$ matrix. Since our structure calculations are performed with the No-Core Shell Model (NCSM) method~\cite{Navratil:2009ut, Barrett:2013nh}, the
three-nucleon ($3N$) interaction is also included in the calculation of the density and our results are obtained for specific choices of the harmonic oscillator frequency
$\hbar \omega$ and $N_{\mathrm{max}}$ parameter, which specifies the number of nucleon excitations above the lowest energy configuration allowed by the
Pauli principle. For the density calculations we also employed the Similarity Renormalization Group (SRG) \cite{Bogner:2003wn, Bogner:2009bt} procedure
to the $NN$ + $3N$ interaction to ensure a faster convergence of our calculation. So, a third parameter characterizing our results is the $\lambda_{\mathrm{SRG}}$
cutoff specifying up to which level the interaction has been evolved.

\begin{figure}[t]
\begin{center}
\includegraphics[scale=0.42]{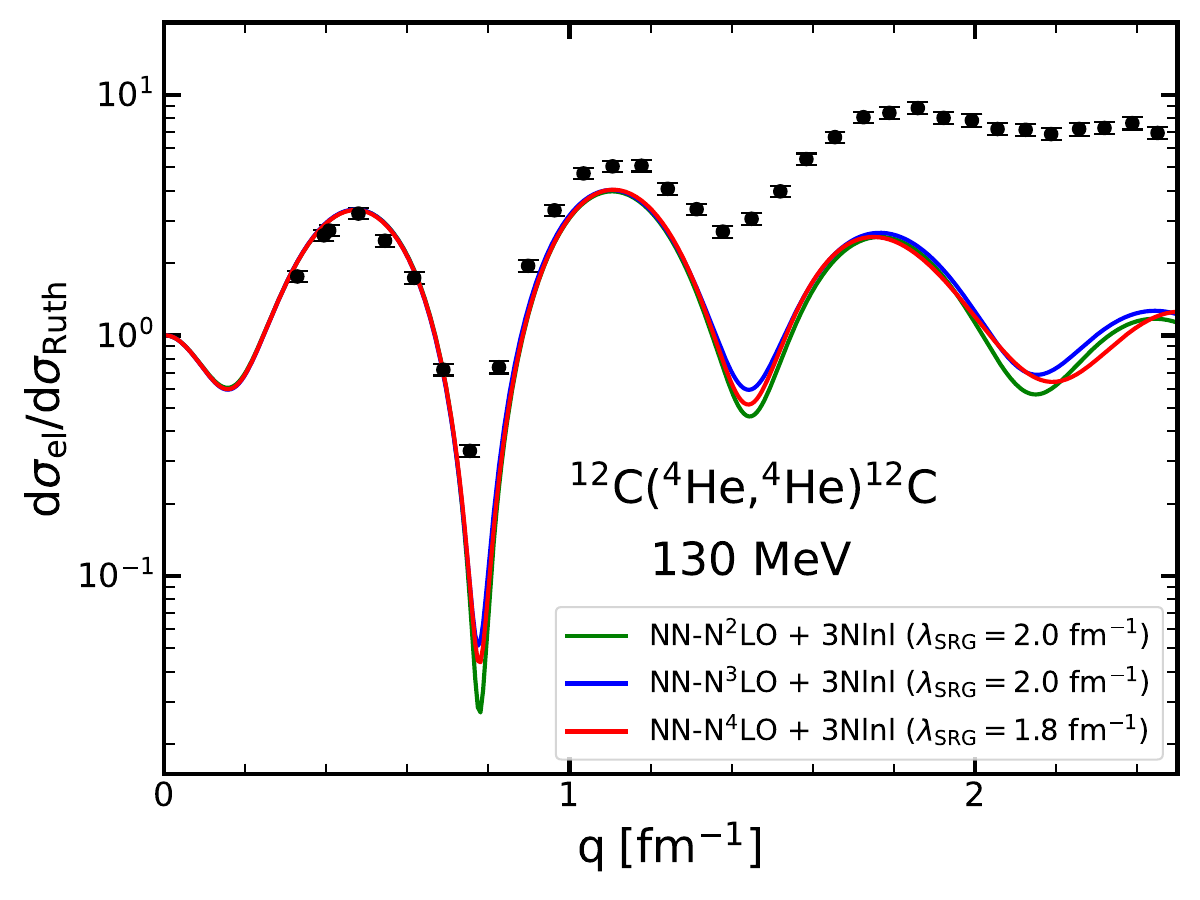}
\caption{
Ratio of the differential cross section to the Rutherford cross section as a function of the transferred momentum $q$ for the reaction $^{12}$C($^{4}$He,$^{4}$He)$^{12}$C
at a projectile energy of 130 MeV. Experimental data \cite{PhysRevC.97.014601} are shown by black circles with corresponding error bars. 
The solid lines are obtained using our microscopic OP of Eq.(\ref{nucleus_opticalpotworkeq_sm}) using different interactions (see text for details) for the calculation
of the $NN$ $t$ matrix and the projectile and target densities.
\label{4He_12C_orders} }
\end{center}
\end{figure}
We start presenting in Figure~\ref{4He_12C_orders} the results obtained with the $NN$ chiral interactions developed by Entem {\it et al.} \cite{PhysRevC.96.024004}
up to the fifth order of the chiral expansion and with a 500 MeV cutoff.
The calculation of the densities is performed with the interactions developed up to the third (N$^{2}$LO), fourth (N$^{3}$LO), and fifth order (N$^{4}$LO) in the $NN$ sector,
supplemented with the $3N$ local-nonlocal chiral interaction at N$^{2}$LO (3Nlnl) presented in Refs. \cite{Navratil:2007zn,PhysRevC.101.014318}.
This interaction has a nonlocal cutoff of 500 MeV and an additional local cutoff of 650 MeV \cite{PhysRevC.101.014318}.
The calculation of the $NN$ $t$ matrix is performed with only the $NN$ part of the interaction.
The low-energy constants $c_D$ and $c_E$ in the $3N$ sector have been fitted to reproduce the triton properties for each choice of the $NN$ interaction. 
At N$^{4}$LO the low-energy constants are fixed at $c_D = -1.8$ and $c_E =-0.31$ \cite{Gysbers2019} while at lower orders we used the values provided
in Table I of Ref.\cite{PhysRevC.102.024616}. The calculation of the densities have been performed with $\hbar \omega = 18$ MeV for both nuclei and
$N_{\mathrm{max}} = 16$ for $^4$He and $N_{\mathrm{max}} = 8$ for $^{12}$C. At N$^{4}$LO the interaction has been evolved up to $\lambda_{\mathrm{SRG}} = 1.8$
fm$^{-1}$ while at lower orders we used $\lambda_{\mathrm{SRG}} = 2.0$ fm$^{-1}$.
We refer to these interactions as NN-N$^n$LO+3Nlnl where $n$ is the order of the expansion in the $NN$ sector. 
As showed in Figure~\ref{4He_12C_orders} the results obtained with these different interactions give very similar results with only minor differences in correspondence
of the diffraction minima and at larger values of momentum transfer.

\begin{figure}[t]
\begin{center}
\includegraphics[scale=0.42]{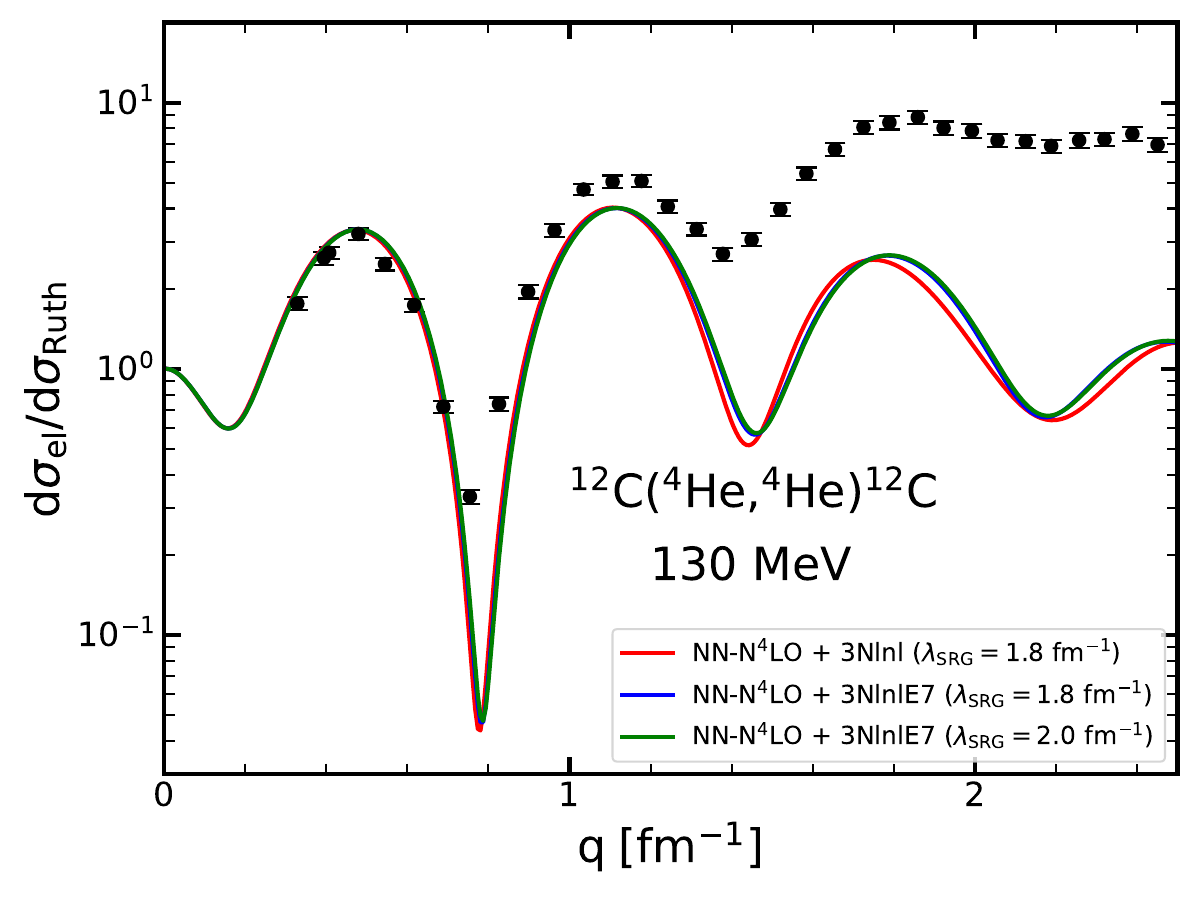}
\caption{
Same as Figure~\ref{4He_12C_orders} but for different choices of interactions and $\lambda_{\mathrm{SRG}}$ (see text for details).
\label{4He_12C_interactions} }
\end{center}
\end{figure}
In Figure~\ref{4He_12C_interactions} we present other results for $^{4}$He elastic scattering off $^{12}$C obtained with other interactions and
for the same projectile energy of 130 MeV.
Since we explored the performance of our OP at different orders of the chiral expansion in the $NN$ sector, now we keep this part
of the interaction constant at N$^4$LO and we use a different $3N$ interaction obtained from the one used in Figure~\ref{4He_12C_orders} where an additional
sub-leading contact term (E7) enhancing the spin-orbit strength \cite{Girlanda2011} has been introduced to the $3N$ force.
The E7 low-energy constant has been adjusted to improve the description of the excitation energies of $^6$Li, in particular of the first excited state $(3^+ , T = 0)$.
We refer to this interaction as NN-N$^4$LO + 3NlnlE7 \cite{Kravvaris2023,PhysRevC.109.065501}. For this interaction we performed calculations
for two different values of the SRG cutoff, namely, $\lambda_{\mathrm{SRG}} = 1.8$ and $\lambda_{\mathrm{SRG}} = 2.0$ fm$^{-1}$.
These results are compared with those obtained with the NN-N$^4$LO + 3Nlnl presented in Figure~\ref{4He_12C_orders}.
Also in this case we obtained very similar results with small differences mostly due to the different $3N$ interaction,
while the difference due to the different choice of $\lambda_{\mathrm{SRG}}$ is practically negligible.
The used microscopic $NN$+$3N$ interactions provide a very good description of the structure of the target nuclei as shown in Ref.~\cite{PhysRevC.109.065501},
where the same interactions were utilized, see Figs. 3 and 4 and Table IV in that paper.

The results presented above justify the choice of N$^4$LO + 3Nlnl interaction with $\lambda_{\mathrm{SRG}} = 1.8$ fm$^{-1}$ as the reference interaction to be
used in the main work.



\nocite{*}

\bibliography{biblio}

\end{document}